\documentclass[12pt]{iopart}
\renewcommand{\d}{{d}} 
\newcommand{\turn}{W}

\usepackage{graphicx}

\usepackage{iopams}
\let\oldref\ref
\renewcommand{\ref}[1]{(\oldref{#1})}

\begin{document}
\title{Stochastic clonal dynamics and genetic turnover in exponentially growing populations}
\author{Arman Angaji, Christoph Velling\footnote{current address: Kishony Lab, Faculty of Biology, Technion City, Haifa 32000, Israel} and Johannes Berg}
\ead{aangaji1@uni-koeln.de,christoph@campus.technion.ac.il, and bergj@uni-koeln.de}
\address{Institute for Biological Physics, University of Cologne,
  Z\"ulpicher Stra{\ss}e 77, 50937 Cologne, Germany}

\vspace{10pt}
\begin{indented}
\item[]June 2021
\end{indented}

\begin{abstract}
We consider an exponentially growing population of cells undergoing mutations and ask about the effect of reproductive fluctuations (genetic drift) on its long-term evolution.
We combine first step analysis with the stochastic dynamics of a birth-death process to analytically calculate the probability that the parent of a given genotype will go extinct. We compare the results with numerical simulations and show how this turnover of genetic clones can be used to infer the rates underlying the population dynamics. Our work is motivated by growing populations of tumour cells, the epidemic spread of viruses, and bacterial growth.
\end{abstract}

%
\vspace{2pc}


\section{Introduction}

Stochastic fluctuations of reproductive success are a key element of evolution. Even under neutral evolution, when on average all clones grow at the same rate, random fluctuations in the number of offspring can drive a clone to extinction. The technical term for such fluctuations in the number of individuals of a particular genotype is genetic drift. Genetic drift affects how a population evolves when mutations conferring higher reproductive rates enter the population (response to selection) and it influences the steady state between clone loss due to reproductive fluctuations and the emergence of new clones through mutations (mutation-drift balance). These effects are very well understood in models with a constant population size, like the Moran model or the Wright-Fisher  model~\cite{kimura1955solution,gillespie2004population,barton2007evolution}. They are far less well understood in populations whose size changes with time.

In this paper, we look at the effects of genetic drift in exponentially growing populations. The   surge in tumour genomic data over the past decade, experiments on growing bacterial populations, and data from the current SARS-CoV-2 pandemic have motivated a wide range of theoretical and bioinformatic studies on populations which grow approximately exponentially~\cite{michor2004dynamics,hallatschek2007genetic,durrett2013population,foo2014evolution,sottoriva2015big,williams2016identification,williams2018quantification,malikic2019integrative,avanzini2019cancer,gunnarsson2021exact,lindstrom2021stochastic}. In an exponentially growing population, neutral evolution leads to expansion of all clones on average, and there are situations when genetic drift only leads to small frequency fluctuations, but not to the extinction of a clone. An example is when individuals reproduce at some rate, but the rate of death can be neglected. At a finite rate of cell death, however, fluctuations can cause a clone to go extinct even when \textit{on average} all clones expand.

In the following, we set up a stochastic framework to describe fluctuations in growing populations. Our focus is on observables at one particular time point (rather than observables which compare the population at two or more different time points). Hence the extinction of one particular clone can only be observed if it has an effect on another clone that has not died out. Specifically, we calculate (i) the probability that the parent of a particular clone dies out and (ii) the analogous quantity for a clade (defined below). The motivation for looking at these particular quantities is twofold. First, they quantify the most drastic consequence of reproductive fluctuations, namely that a clone can die out. The second reason is practical: the standard observable of population genetics, the frequency spectrum of mutants, depends on the mutation rate $\mu$ and the rates of cell birth $a$ and death $b$ via $\mu a/(a-b)$~\cite{durrett2013population} (at least for frequencies which are not very small, see~\cite{gunnarsson2021exact}). A change in $\mu$ and a change in $b$ can have the same effect on the frequency spectrum. For a separate inference of the parameters underlying the population dynamics we thus need to look beyond the frequency distribution. It turns out that the parameters describing the turnover of clones and clades allow to infer the parameters underlying the population dynamics.

Useful tools and concepts beyond the birth-death process~\cite{novozhilov2006biological} turn out to be first step analysis~\cite{privault2013understanding,durrett2015branching} to describe the extinction probability of a clone and the exponential distribution over the time period which mutations have between the moment they arise and a final time point when observables are taken.

In principle, the growth rate of an individual can depend on its position in space, and this dependence can be different for different systems. In an extreme case, growth is restricted to the surface of a growing population, rather than occurring uniformly across its bulk, and the resulting reduced number of growing cells leads to a large genetic drift~\cite{hallatschek2007genetic}. In two-dimensional bacterial colonies, this population edge is one-dimensional, and has been analyzed using the linear stepping stone model~\cite{korolev2010genetic}. In tumours, cell dispersion near the edge of the tumour has been put forward as an important driver of evolution~\cite{waclaw2015spatial}, although it is not clear how much spatial effects influence the genetic record. Epidemics spread both locally and along the long-range networks of air travel~\cite{hufnagel2004forecast}, and thus also involve spatial effects. In this paper, however, we focus on the constant-rate birth-death process and leave spatial effects for future work.

\section{Stochastic dynamics of clones and clades}


We consider an asexually growing population of cells (or other entities) dividing at fixed rate
$a$ per cell and dying at fixed rate $b$. Upon cell division,
each offspring can either gain a mutation (an event that happens with probability $\mu$), or not gain a mutation
(probability $1-\mu$). The mutations of the different offspring occur independently, and mutations
can only be gained but not lost (no backmutations). The mutations arising in different events
are distinguishable from one another, and can be labelled in the order of their occurrence. Such a scenario arises in the limit of an infinite genome (infinite sites model), where every new mutation occurs on a novel site, and at low mutation rates (so the cases of multiple mutations per division can be neglected). Low mutation rates can be implemented in practice by considering sufficiently small parts of the genome, see below.

A particular set of mutations defines a genotype, and the cells with a particular genotype define a clone. The size of the clone varies with time through the stochastic birth and death of cells, as well as through further mutations (since every new mutation introduces a new genotype and hence a new clone). Analogously to the dynamics of cells sharing the same genotype, we can also look at the set of cells sharing a particular mutation. The cells sharing a particular mutation form a so-called clade, which is defined by that mutation. The size of a clade also varies with time through the birth and death of cells, but not through further mutations.

Both the dynamics of a clone and a clade can be discussed within the same framework by defining effective rates $\alpha$ and a death rate $\beta$ at which these sub-populations grow and shrink. For the cells descending from a particular mutant cell (the clade) they are simply
\begin{eqnarray}
\label{eq:ratesclades}
\alpha&=&a  \\
\beta &=&b . \nonumber
\end{eqnarray}
For a clone they are given by
\begin{eqnarray}
\label{eq:ratesclones}
\alpha&=&a(1-\mu^2)  \\
\beta &=&b + a\mu^2 ,\nonumber
\end{eqnarray}
which are the rate at which cells divide and neither offspring mutates (increasing the size of a clone), and the rate at which cells either die or divide with mutations occurring in both offspring (decreasing the size of a clone). From now on, we will refer to the general rates $\alpha$ and $\beta$, and for simplicity refer to the dynamics of a clone, although using either ~\ref{eq:ratesclades} or~\ref{eq:ratesclones} the dynamics of a clade or of a clone can be described.

\begin{figure*}[tb!]
A
\includegraphics[width=0.4\textwidth]{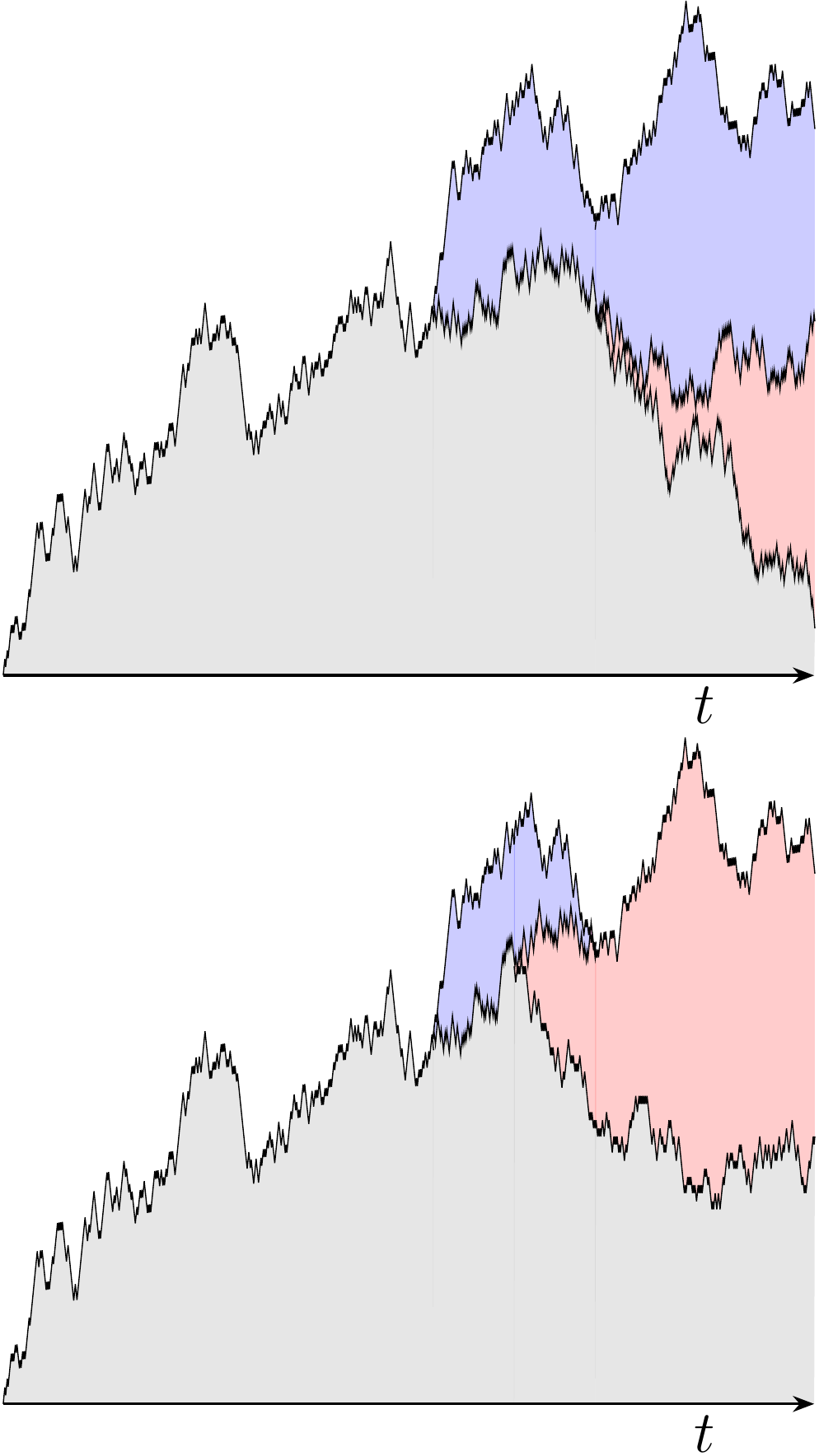}
\hfill
B
\includegraphics[width=0.4\textwidth]{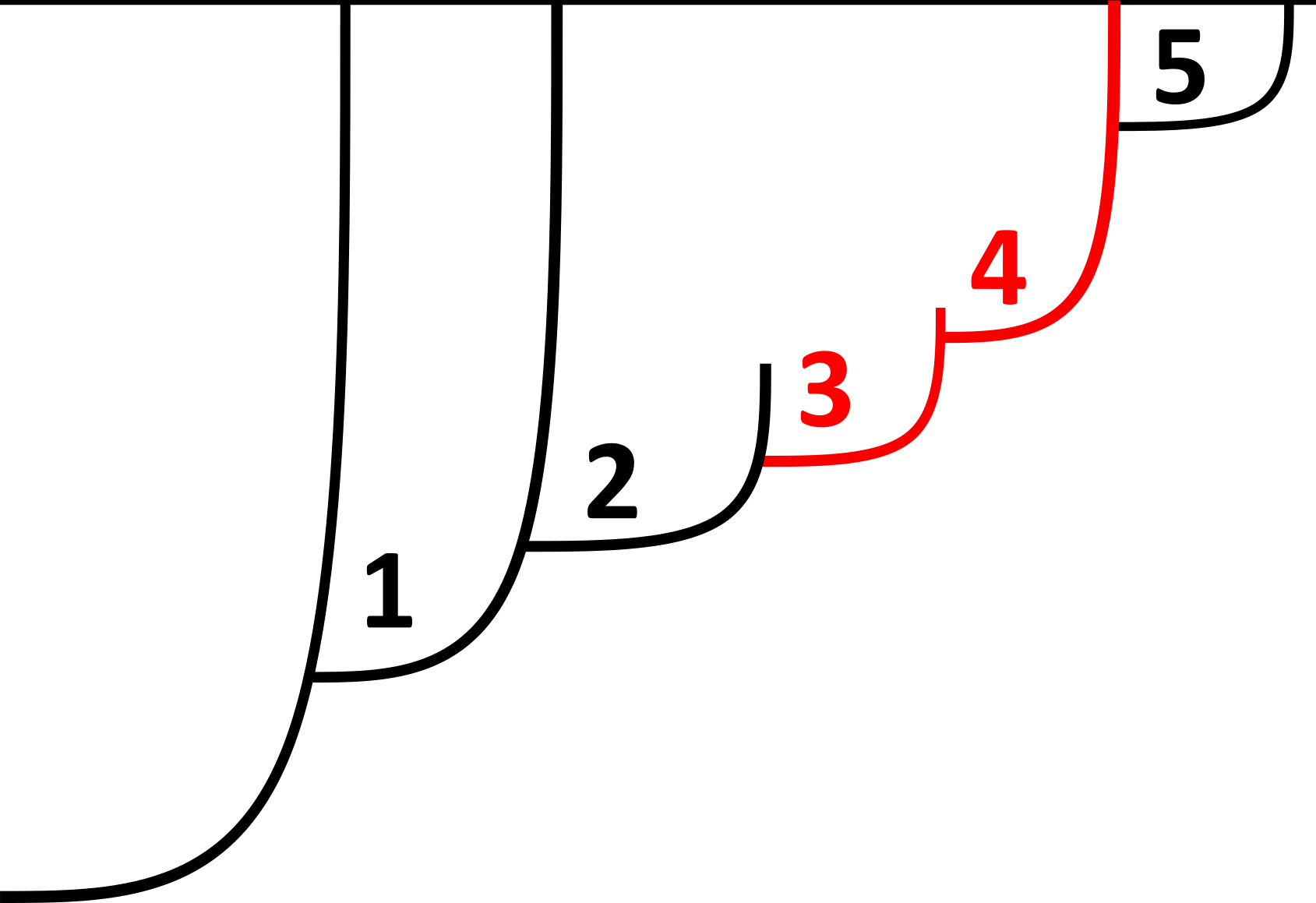}
\caption{A: A schematic illustration of the growth of two populations, in which a mutant (blue) arises. On the mutant background, a further mutation, shown in red, emerges. In the case shown on top, the parental (blue) clone survives, in the case shown on the bottom the parental clone dies out. The number of the parental clones going extinct relative to the total number of clones specifies the clone turnover $W_o$. The same figure can also be though to represent clades; the red clade that is a sub-clade of and part of the blue clade. The lower figure then shows an example where the new (red) mutation has become clonal in the parental clade. The fraction of these
cases specifies the clade turnover $W_a$.
B: An example of a phylogenetic tree where some branches have become extinct to illustrate the computation of the turnover parameters, see text. The clones indicated with mutations $3$ and $4$ lack an ancestral clone and thus contribute to both the clone turnover and the clade turnover.}
\label{fig:clonesschematic}
\end{figure*}

We consider a population that starts from a single cell and grows for a time $T$, see Fig. \oldref{fig:clonesschematic}A. Since clones are born at different times, each clone has a different period during which its dynamics runs from its birth to the final time $T$. In an exponentially growing population, most clones are born close to the final time, since then the population size is largest. The distribution of these "running times" is exponential; picking cells uniformly from all the cells that have been born, the probability a cell was born at time $t$ is proportional to the population size $N(t)=e^{(a-b)t}\equiv e^{\gamma t}$ at the time. $\gamma \equiv (a-b)$ is a shorthand for the population growth rate. The normalizing factor is $(e^{\gamma T}-1)/\gamma$, yielding for large final times $T$ the distribution $\gamma e^{\gamma (t-T)}$, an exponential distribution for the running times $\tau \equiv T-t$.

We now consider a clone that grows stochastically or an ensemble of such clones. For convenience we set the time of birth to $t=0$, and model its running time as an exponentially distributed random variable. The stochastic dynamics of the size $n(t)$ at subsequent times is described by the straightforward master equation
\begin{eqnarray}
\frac{d}{dt}p_n(t)&=&\alpha (n-1) p_{n-1}(t) - (\alpha+\beta)n p_n(t) \\
&&+ \beta (n+1) p_{n+1}(t)
-\gamma p_n(t) \ . \nonumber
\label{eq:master}
\end{eqnarray}
The first part of this equation describes a standard birth-death process~\cite{privault2013understanding} and has been used extensively to describe growing populations~\cite{durrett2015branching}. The final term, $-\gamma p_n(t)$, describes the exponential decay of probability due to the finite running time.

The standard approach to such master equations is to define the generating function
\begin{equation}
z(q,t)=\sum_{n=0}^{\infty} q^n p(n,t)
\end{equation}
for which the master equation~\ref{eq:master} yields the dynamical equation
\begin{equation}
\frac{\partial}{\partial t}z(q,t)= (\alpha q^2 - (\alpha+\beta)q + \beta)
  \frac{\partial}{\partial q}z(q,t) - \gamma z(q,t) \ .
  \label{eq:zdyn}
\end{equation}
for the generating function $z(q,t)$.

The boundary condition for this linear partial differential equation with constant coefficients is $z(q,t=0)=q$ since at time $t=0$ the clone consists of a single cell, so $p_n(t=0)=\delta_{n,1}$. Its solution is
\begin{equation}
z(q,t)= e^{-\gamma t}
\frac{(1-q)\beta-e^{-(\alpha-\beta) t}(\beta-\alpha q) }
{(1-q)\alpha-e^{-(\alpha-\beta) t}(\beta-\alpha q) } \ .
  \label{eq:zsol}
\end{equation}
This solution of the generating function allows in principle to derive the statistics of clone
sizes at all times by taking derivatives of the generating function with respect to $q$ to determine $p_n(t)$.

A second, less standard use of the generating function arises in the context of clones becoming extinct. To compute the probability that a given cell \textit{and} all of its offspring will eventually die we use the so-called first step analysis~\cite{privault2013understanding,durrett2015branching}. This elegant and straightforward method considers a particular cell at its birth, and denotes the probability that the cell and its offspring will go extinct by $q$. An arbitrary small time-step $\Delta t$ later, there are three possible outcomes: with probability $p_0=\beta \Delta t$ the cell has died, with probability $p_1=1-(\alpha+\beta)\Delta t$ it has neither died nor divided, and with probability $p_2=\alpha \Delta t$ the cell has divided. In order for the cell and its descendents to die, the event has already happened in the first case, one cell and its descendents need to die in the second, and two cells and their descendents need to die in the third. In the last case, the death of each of the two cells (and their offspring) are statistically independent; the event thus occurs with probability $q^2$. This gives a self-consistent equation for $q$
\begin{equation}
p_0+p_1 q + p_2 q^2 = q \ ,
\end{equation}
which is solved for arbitrary small $\Delta t$ by $q=\beta/\alpha$.

The link with the generating function arises when we consider a clone that is growing stochastically and at time time $t$ consists of $n$ cells with probability $p_n(t)$. The probability that this clone will die out is given by $\sum_{n=0}^{\infty} (\beta/\alpha)^n p(n,t)=z(\beta/\alpha,t)$.

\section{Clade and clone turnover}

We now combine the dynamics of the generating function $z(q,t)$ with the result of the first step analysis in order to address the following question: What is the probability that the parent of some clone dies out? The probability of a clone itself dying out is simple, and is given by the first step analysis above. However, the death of a clone is not directly observable from extant cells (cells present at time $T$) at a later time as there is no evidence that the clone ever arose. If, on the other hand, one can identify the parent of a clone that parent is either extant in the population, or it is not. One can then compare the empirical fraction of clones without a parent with the probability that the parent of a clone has been lost. Clones can be lost by cell death or by mutations as specified by the rates \ref{eq:ratesclones}. A parental clade, by comparison, is lost through cell death only. In that case, the parental clade is replaced by the offspring (see Fig.~\oldref{fig:clonesschematic}A); the offspring has become what is called \emph{clonal} in the parental clade.

The probability that a clone emerges from some parental clone of size $n$ is proportional to $n p_n(t)$, and the probability that the parental clone dies out is given by $(\beta/\alpha)^n$. Denoting again the time at which the parental clone arose by $t=0$, and averaging over all the times and parental clone sizes at which the offspring might have arisen we obtain the \textit{turnover parameter}
\begin{equation}
\turn=\frac{\int_0^T \d t \sum_{n=0}^{\infty} n p_n(t) (\alpha/\beta)^n}
     {\int_0^T \d t \sum_{n=0}^{\infty} n p_n(t) }
     =
  \frac{q\int_0^T \d t \partial_q |_{q=\beta/\alpha} z(q,t)}
          {\int_0^T \d t \partial_q |_{q=1}z(q,t) }
\label{eq:turnover}
\end{equation}
which gives the expected number of clones whose parental clones have died out due to fluctuations in
the number of cell births, deaths, and (for genotypes) mutations. It is a measure of how fast clones are lost in the population due to genetic drift.

To evaluate this expression we consider its numerator and denominator separately and define
\begin{equation}
G(q,T)=\int_0^T \d t \sum_{n=0}^{\infty} n p_n(t) q^n = q\int_0^T \d t \partial_q z(q,t) \ ,
\label{eq:Gdef}
\end{equation}
where $G(1,T)$ gives the denominator and $G(\beta/\alpha,T)$ gives the numerator of the turnover.
Since the partial differential equation~\ref{eq:zdyn} and $G(q,T)$ are both linear in the generating function $z(q,t)$ the calculations to determine the turnover are well handled by a computer algebra system. From~\ref{eq:zsol} we obtain
\begin{eqnarray}
  \label{eq:Geval}
G(q,T)&=&
\left[
\alpha(1-q)-\frac{
\gamma(\alpha-\beta)\ _2 F_1(2,1+\frac{\gamma}{\alpha-\beta},2+\frac{\gamma}{\alpha-\beta},\frac{\beta-\alpha q}{\alpha-\alpha q})}
{\alpha-\beta+\gamma}
\right.\\
&&\left.
+(\alpha-\beta)\e^{-(\alpha-\beta+\gamma)T}
\left(
-\frac{\alpha(1-q)}{\alpha(1-q)+(q\alpha-\beta)\exp^{-(\alpha-\beta)T}}
\right.\right.
\nonumber\\
&&\left.\left.
+\frac{
\gamma \ _2 F_1(2,1+\frac{\gamma}{\alpha-\beta},2+\frac{\gamma}{\alpha-\beta},\e^{-(\alpha-\beta)T}\frac{\beta-\alpha q}{\alpha-\alpha q})
}
{\alpha-\beta+\gamma}
\right)
\right]/(\alpha^2(1-q)^2)
\nonumber
\end{eqnarray}
where $_2 F_1$ is the Gaussian hypergeometric function.
For the denominator of the turnover parameter we obtain from the asymptotic limit when the last argument of the hypergeometric function is taken to infinity
\begin{equation}
G(q=1,T)=\frac{1-e^{(\alpha-\beta-\gamma)T}}{\gamma-(\alpha-\beta)} \ .
\label{eq:G1}
\end{equation}
For the numerator of the turnover parameter we obtain using $_2 F_1(\ldots,\ldots,\ldots,\frac{\beta-\alpha q}{\alpha-\alpha q})=1$ for $q=\beta/\alpha$
\begin{equation}
G(q=\beta/\alpha,T)=\frac{1-e^{-(\alpha-\beta+\gamma)T}}{\gamma+(\alpha-\beta)}
\label{eq:Gbetaoveralpha}
\end{equation}
giving the turnover parameter~\ref{eq:turnover} as
\begin{equation}
\turn= \frac{\beta}{\alpha}
    \frac{\gamma-(\alpha-\beta)}{\gamma+(\alpha-\beta)}
    \frac{1-e^{-(\alpha-\beta+\gamma)T}}{1-e^{(\alpha-\beta-\gamma)T}} \ .
\label{eq:turnover2}
\end{equation}

The next step is to evaluate the turnover parameter separately for the clade defined by a mutation using the rates \ref{eq:ratesclades} and for a clone using rates \ref{eq:ratesclones}. For the descendants of a mutant, we insert the rates \ref{eq:ratesclades} into the turnover
\ref{eq:turnover2} and take the limit $\gamma \to a-b$ which gives the \textit{clade turnover}
\begin{equation}
\turn_a=  \frac{b/a}{2(a-b)T} (1-e^{-2(a-b)T}) \ .
\label{eq:turnover_clades}
\end{equation}
For the turnover of clones we insert the rates \ref{eq:ratesclones} into the turnover
\ref{eq:turnover2} and obtain the \textit{clone turnover}
\begin{equation}
\turn_o=  \frac{\mu(b+a\mu^2)}{(1-\mu)^2(a-b-a\mu)} \frac{1-e^{-2(a-b-a\mu)T}}{1-e^{-2a\mu T}}  \ .
\label{eq:turnover_clones}
\end{equation}
For the clone turnover the result depends on the final time $T$ only exponentially weakly; for large times $T$, the second factor in~\ref{eq:turnover_clones} asymptotically tends to one (for times large relative to the inverse rates of division and death), so the turnover becomes  independent of $T$. This is different for the clade turnover~\ref{eq:turnover_clades}, which always depends on final time $T$, which appears in the denominator of ~\ref{eq:turnover_clades}. The reason for this dependence is that as the clade defined by a particular mutation grows at the same rate as the rest of the population, the expected frequency of that mutation remains fixed at $1/N(t)$, the inverse population size when it arose~\cite{williams2016identification}. Thus even old mutations contribute to the turnover and the clade turnover depends strongly on the upper limit of the integral over time in the turnover~\ref{eq:turnover}. Conversely, the rate at which a clone grows is diminished by mutations (which generate new clones), and hence clones grow more slowly than the population size.  As a result, for the clone turnover, old mutations contribute only exponentially weakly to the integral over time in ~\ref{eq:turnover}.

Both the clone and the clade turnover parameter have straightforward interpretations. The clone turnover gives the fraction of genotypes whose parental genotype is no longer extant. A schematic example is given in Fig. \oldref{fig:clonesschematic}A. The clade turnover gives (averaged over the different clades) the fraction of clades that coincide with their ancestral clade, i.e. the novel mutation has become clonal in the parental clade.

These observables are best illustrated with a concrete example. We consider a population in which mutations $1, 2, 3, 4, 5$ arose in that order
(the order is given only to illustrate the example, see Fig. \oldref{fig:clonesschematic}B).
In this population, clones carrying mutations $\{\}$, $\{1\}$, $\{1,2,3,4\}$, $\{1,2,3,4,5\}$ survived.
Clone $\{1\}$ has an extant parental genotype, namely $\{\}$. None of the extant genotypes carry exactly three mutations out of $\{1,2,3,4\}$. The parental clone of this genotype thus became extinct. Finally genotype $\{1,2,3,4,5\}$ has an extant parent in $\{1,2,3,4\}$. Hence $2$ out of $3$ clones have extant parents, resulting in a clone turnover of $\turn_o = 2/3$.

In order to compute the clade turnover we go over all mutations $1, 2, 3, 4, 5$, which each define a clade.
The only ancestral clade of mutant $1$ is the original clade which gave birth to all clades. The clone $\{\}$ survived, therefore clade $1$ does not coincide with its ancestor. Mutation $1$ hence contributes one to the denominator and zero to the numerator of the clade turnover. Mutation $2$ does not coincide with either of its two ancestral clades (origin and $1$) and thus adds two to the denominator and zero to the numerator. Mutation $3$ is again not clonal on the original clade and $1$, but it coincides with $2$. Its contribution to the clade turnover is three in the denominator and one in the numerator.
The same goes for the clade of $4$, which in addition coincides with $3$, resulting in a contribution of four to the denominator and two to the numerator. Lastly, clade $5$ has the original clade and mutants $1, 2, 3, 4$ as ancestral clades and coincides with neither of them, adding five to the denominator and zero to the numerator.
The clade turnover is hence $\turn_a = 3/15 = 1/5$.

This direct procedure assumes that we know the order in which the mutations occurred, but in fact the clade turnover can be computed without knowledge of the phylogenetic tree. For a given mutant clade $m$ we first identify the clonal set of mutations (also called the \emph{trunk}; mutations which always co-occur with $m$) and the so-called \emph{private subset of truncal mutations}; those truncal mutations which are unique to that clade.
The ancestors are all represented in the trunk, but $m$ can only be clonal on ancestors which are part of its private subset. We do not know the order in which the mutations in the private subset arose, but we can compute a combined contribution to the clade turnover. Unique trunks contribute $r(r-1)/2 + r (t-r+1)$ and $r(r-1)/2$ to the denominator and the numerator of the clade turnover respectively, where $r$ is the size of the private subset and $t$ the size of the trunk.

In the given example, mutations $2,3,4$ have a common trunk $\{1,2,3,4\}$ of size $t=4$ with $r=3$ private truncal mutations $\{2,3,4\}$. They together add nine to the denominator and three to the numerator. Mutant $1$ has $r=t=1$ and thus contributes zero to the numerator and one to the denominator, mutant $5$ has $t=5$ and $r=1$, contributing zero to the numerator and $5$ to the denominator, yielding the same clade turnover as above.

In order to evaluate both turnover parameters in empirical data, we need to distinguish the mutations which arose in different generations. This can easily be achieved when the mutation rate is low, and typically there is at most one mutation per generation. In practice, this can always be achieved by restricting mutations to a particular part of the genome, see below.

\section{Numerical Simulations}

We used the Gillespie algorithm~\cite{gillespie1977exact} to simulate the exponential growth of mutant clones within a population of cells. To measure the turnover from simulations, a population is first grown to a threshold size $N$. The growth process then continues to a size that is sufficiently high to observe whether a parental clone has gone extinct. In the case of clade turnover, the mutation probability can be set to zero in the second phase, because further mutations do not affect the fate of a clade. The results are averaged over a few hundred runs depending on mutation probability (which increases the number of clones and hence the computational cost).

\begin{figure*}[t!]
\includegraphics[width=\textwidth]{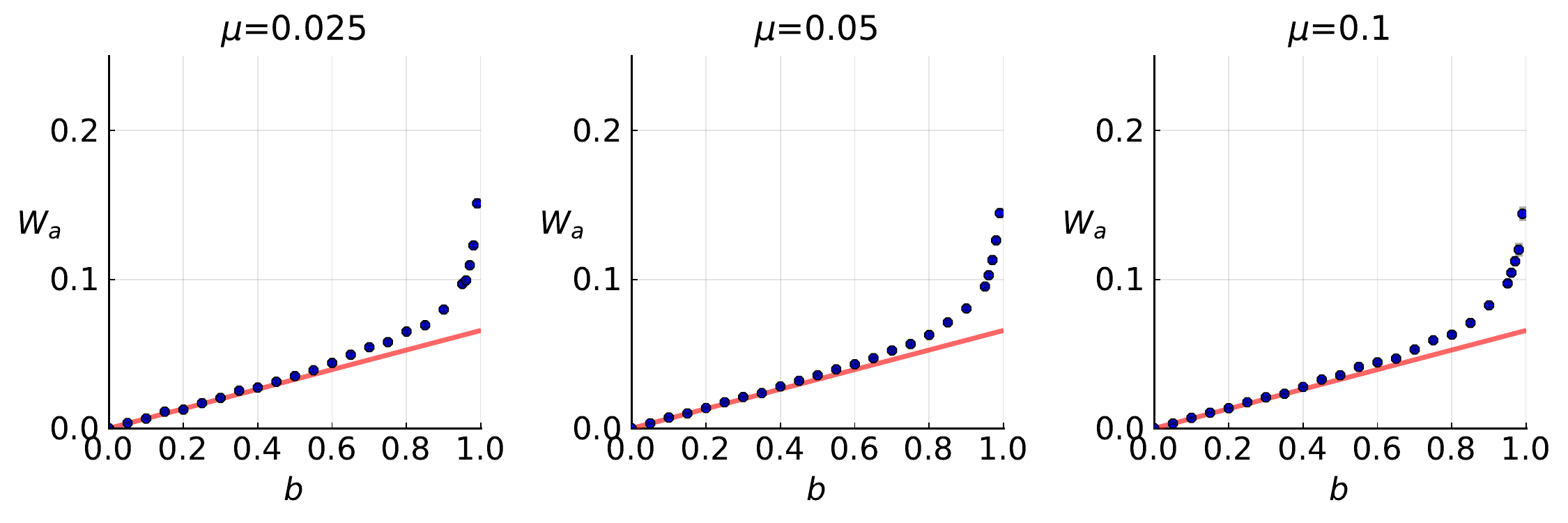}
\caption{The clade turnover calculated from numerical simulations (circles) is shown together with the analytical result \ref{eq:turnover_clades} (solid line) for different mutation probabilities $\mu=0.025,0.05,0.1$ (left to right). Simulations were averaged over $800$, $500$, and $200$ populations, respectively. The birth rate was $a=1$, a threshold of $N=2000$ cells and a final population size of $6000$ cells were used. The standard error corresponds roughly to the size of the symbols.}
\label{fig:cladeturnover}
\end{figure*}

\begin{figure}
\includegraphics[width=\textwidth]{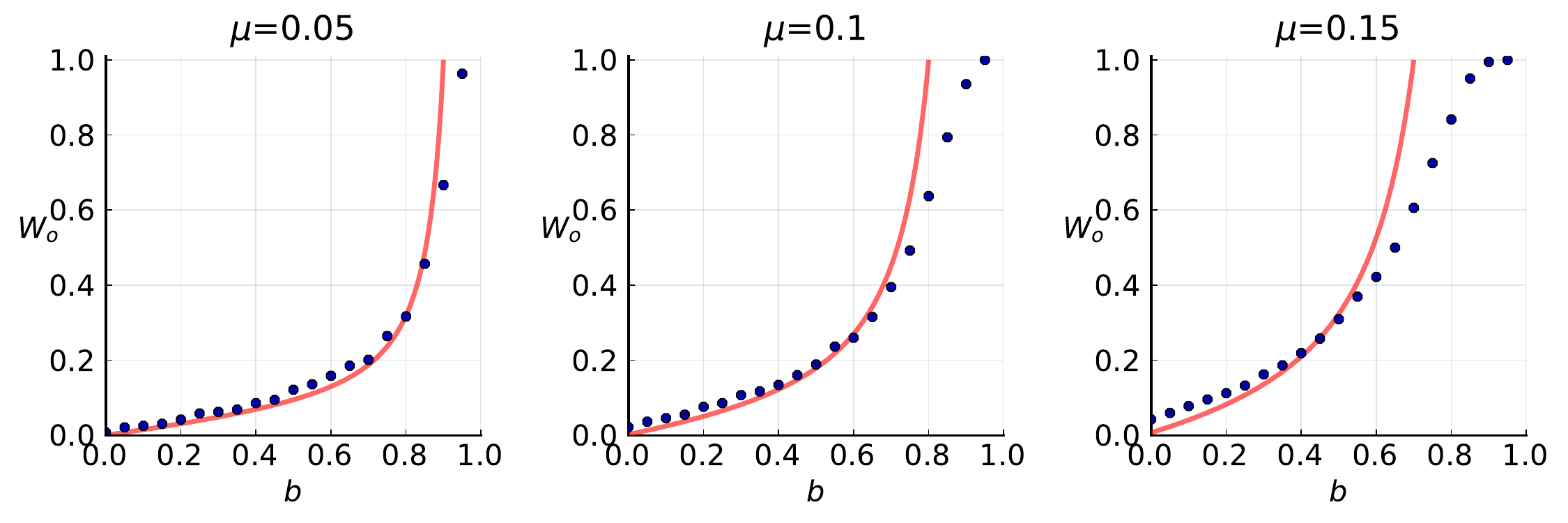}
\caption{The clone turnover calculated from numerical simulations (circles) is shown together with the analytical result \ref{eq:turnover_clones} (solid line) for different mutation probabilities $\mu=0.025,0.05,0.1$ (left to right). Simulations were averaged over $800$, $500$, and $200$ populations, respectively. The birth rate was $a=1$, a threshold of $N=200$ cells and a final population size of $2000$ cells were used. By comparison with Figure~\oldref{fig:cladeturnover}, a smaller population size was used since to compute the clone turnover mutations must be considered also during growth from the threshold size $N$ to the final population size. This increases the computational effort (see text). The standard error corresponds roughly to the size of the symbols.}
\label{fig:cloneturnover}
\end{figure}

Figures \oldref{fig:cladeturnover} and \oldref{fig:cloneturnover} show very good agreement between the numerical simulations and the analytical results \ref{eq:turnover_clades} and \ref{eq:turnover_clones}. Only at high rates of cell death (or high mutation probabilities for the clone turnover, which also increase the effective death rate \ref{eq:ratesclones} of a clone) there are deviations between the numerical and analytical results. These deviations decrease as the threshold $N$ and the times waiting for a line to potentially go extinct are increased. In this sense, the deviations can be viewed as finite-size effects, which lead to the smoothing of discontinuities present in the analytical results: For the clone turnover~\ref{eq:turnover_clones}, there is a kink when the effective birth and death rates
\ref{eq:ratesclones} become equal to one another and the turnover reaches one. The clade turnover jumps discontinuously to one when the birth and death rates \ref{eq:ratesclades} become equal. The clade turnover turns out to be independent of the mutation probability and the corresponding simulation results for different mutation probabilities are identical up to sampling noise. This is because once a clade is arisen, its dynamics is independent of further mutations.

We also considered the number of mutations arising at a division to be distributed as a Poisson random variable with mean $\mu$. The results do not differ considerably at small mutation probabilities. In practice, the mutation probability can be tuned to small probabilities by considering mutations in parts of the genome only, thus reducing the size of the mutational target, an effect we will use for parameter inference below.

\section{Parameter inference}

Given a single population in which the turnover parameters are measured, we ask if the
underlying parameters of the population dynamics can be inferred; specifically the death rate relative to the division rate $b/a$, and the mutation probability $\mu$.

From the clade turnover \ref{eq:turnover_clades}, the ratio between death rate and division rate follows directly as
\begin{equation}
  \label{eq:celldeathfromcladeturnover}
b/a = 2 \turn_a \log(N)/(1-N^{-2}) \approx 2\log(N) \turn_a \ .
\end{equation}
for large $N$. In a next step, given this ratio and the clone turnover $\turn_o$, \ref{eq:turnover_clones} can be solved for $\mu$.

\begin{figure}[tb!]
\includegraphics[width=.75\textwidth]{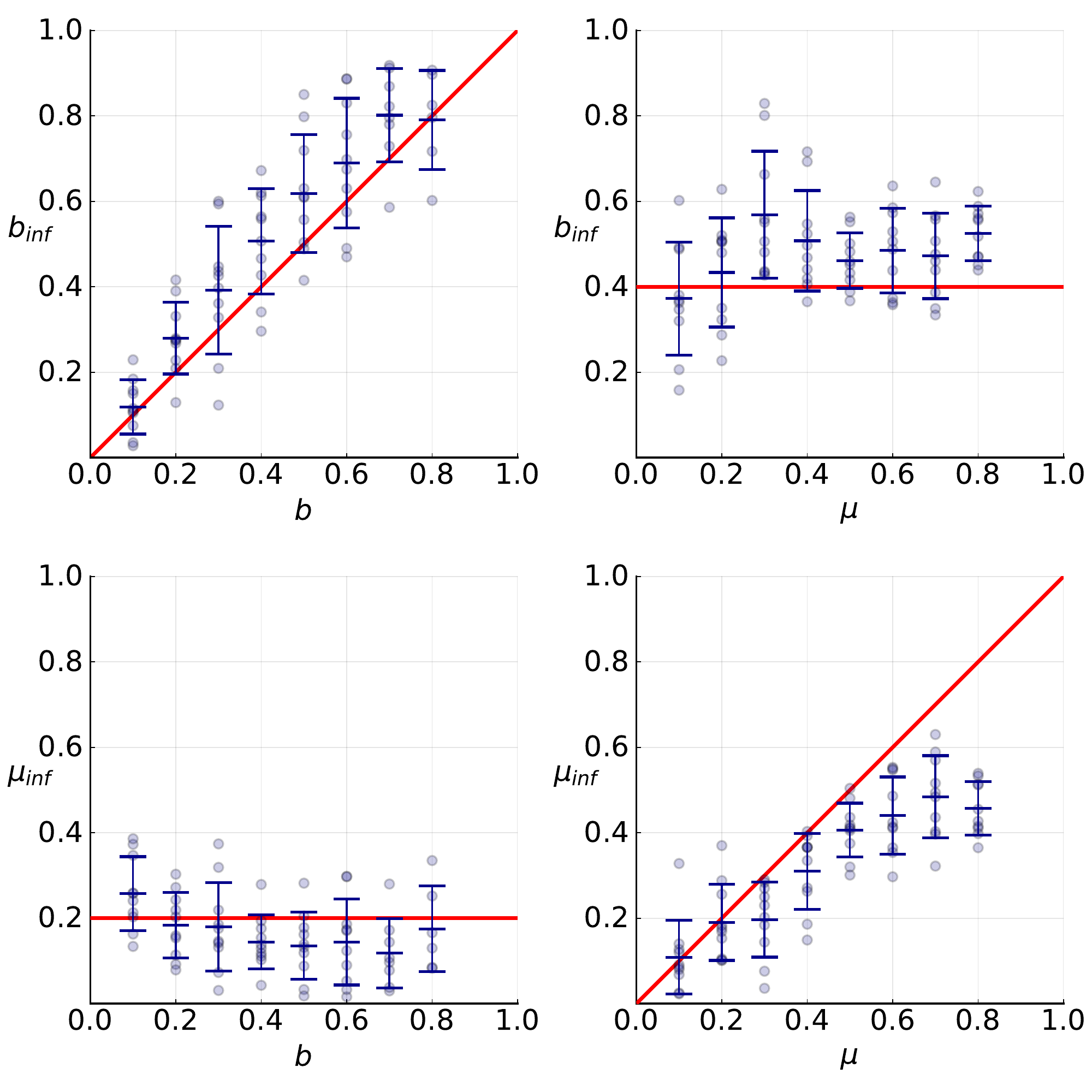}
\caption{The inference of the rate of cell death $b$ (top) and the mutation probability $\mu$
(bottom) by solving the clade turnover \ref{eq:turnover_clades} and the clone turnover
\ref{eq:turnover_clones}. On the left, the death rate is varied at a constant mutation probability $\mu=0.2$, on the right, the mutation probability is varied at a constant death rate of $b=0.4$. The cell division rate is kept constant at $a=1$. The inferred death rate $b_{\textrm{\tiny{inf}}}$ and the inferred mutation probability $\mu_{\textrm{\tiny{inf}}}$ are shown on the $y$-axes of the top and bottom figures, respectively. The solid lines indicates perfect inference. Small circles indicate individual populations,
error bars indicate the mean and standard deviation of the inferred parameters.}
\label{fig:inference1}
\end{figure}

To test this simple inference procedure,
single populations were grown at the specified rates to a size of $10^5$ cells and the clone and clade turnover were computed as detailed above for mutations carried at least by $N=2000$ cells. This was to ensure that there was sufficient time for the parental clade or clone to die out. The inference results for different values of the death rate and the mutation probability are shown in Figure~\oldref{fig:inference1}.

\begin{figure}[tb!]
\includegraphics[width=.75\textwidth]{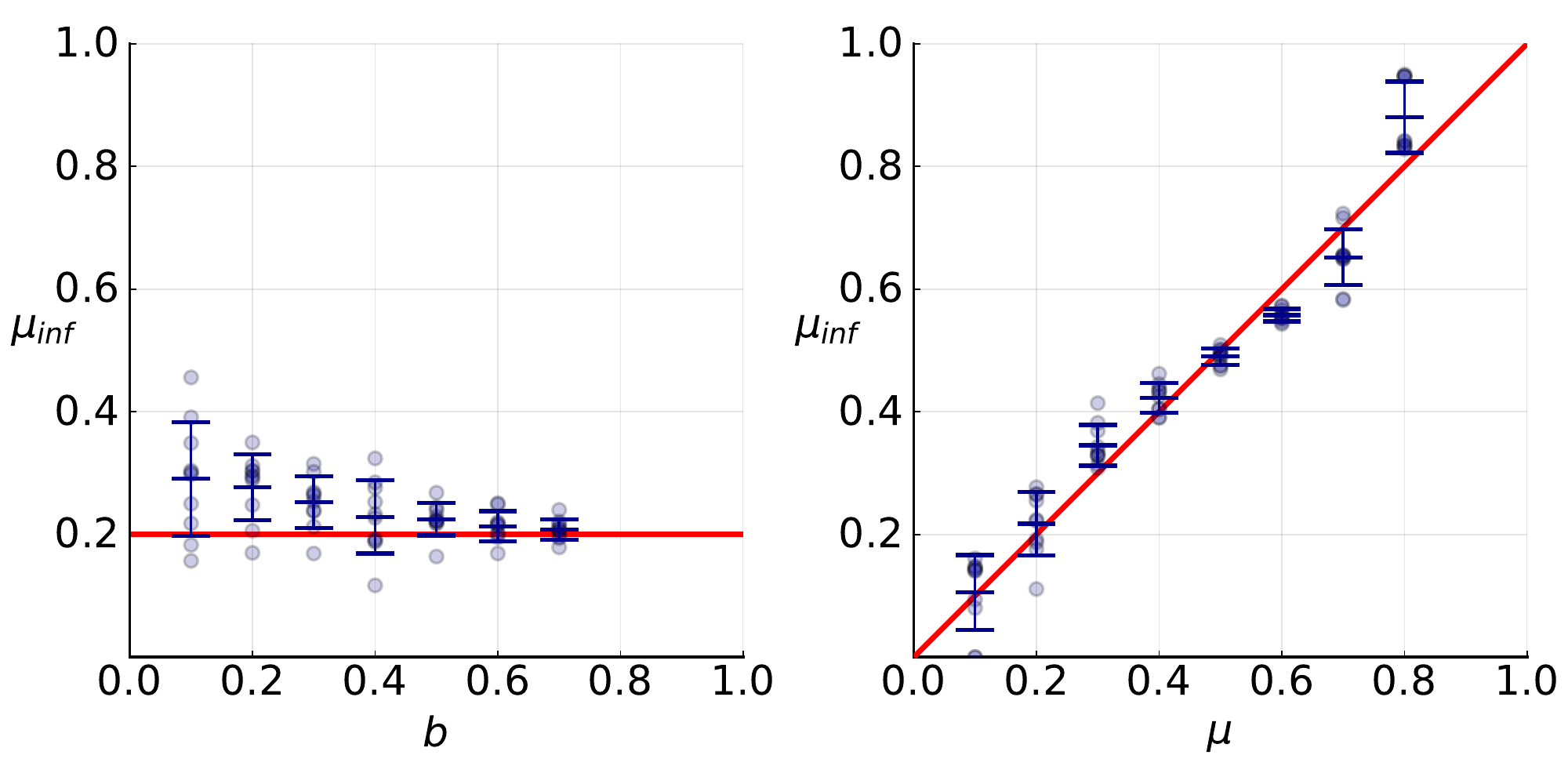}
\caption{The inference of the mutation probability $\mu$
by fitting the clone turnover \ref{eq:turnover_clones} as a function the mutation rate (tuned by considering only a fraction of the genome, see text). On the left, the death rate is varied at a constant mutation probability $\mu=0.2$. On the right, the mutation probability is varied at a constant death rate of $b=0.4$. The remaining simulation parameters are also as in Figure~\oldref{fig:inference1}.
}
\label{fig:inference2}
\end{figure}

Given the relative rate of cell death from~\ref{eq:celldeathfromcladeturnover}, the mutation probability can also be inferred by a fit to the clone turnover~\ref{eq:turnover_clones}. We determine $b/a$ from \ref{eq:celldeathfromcladeturnover} as before. By considering mutations
only from an (arbitrarily chosen) fraction $\delta$ of the genome, the mutation probability $\mu \delta$ can be controlled by changing the fraction $\delta$.
Inserting $\mu \delta$ for the mutation probability in~\ref{eq:turnover_clones}, $\mu$ can then be inferred by fitting the observed clone turnover as a function of $\delta$ to \ref{eq:turnover_clones} using $aT=\log(N)/(1-b/a)$. The results are shown in Figure~\oldref{fig:inference2}. Both methods give a good reconstruction of the model parameters on the basis of a single population.

So far, we have assumed perfect sampling of the population. In practice, however, a population is rarely available entirely for analysis. Instead, many datasets are generated from a limited number of samples taken from the population, for instance when parts of a tumour are excised in a biopsy. Clearly our method will fail if the presence or absence of parental clones cannot be determined due to a low sampling rate. In principle, the effects of sampling can be corrected for by including sampling into a statistical model of the population dynamics~\cite{stadler2009incomplete}. Here we illustrate how our method can be applied to current data with finite sampling in the context of cancer. However, we expect that the sampling problem will become largely irrelevant in that context with the advent of whole-tumour single-cell data, where the genotype of a cell is determined for thousands of cells sampled uniformly from a whole-tumour resection.

\begin{figure}[tbh]
    \includegraphics[width = 0.75\textwidth]{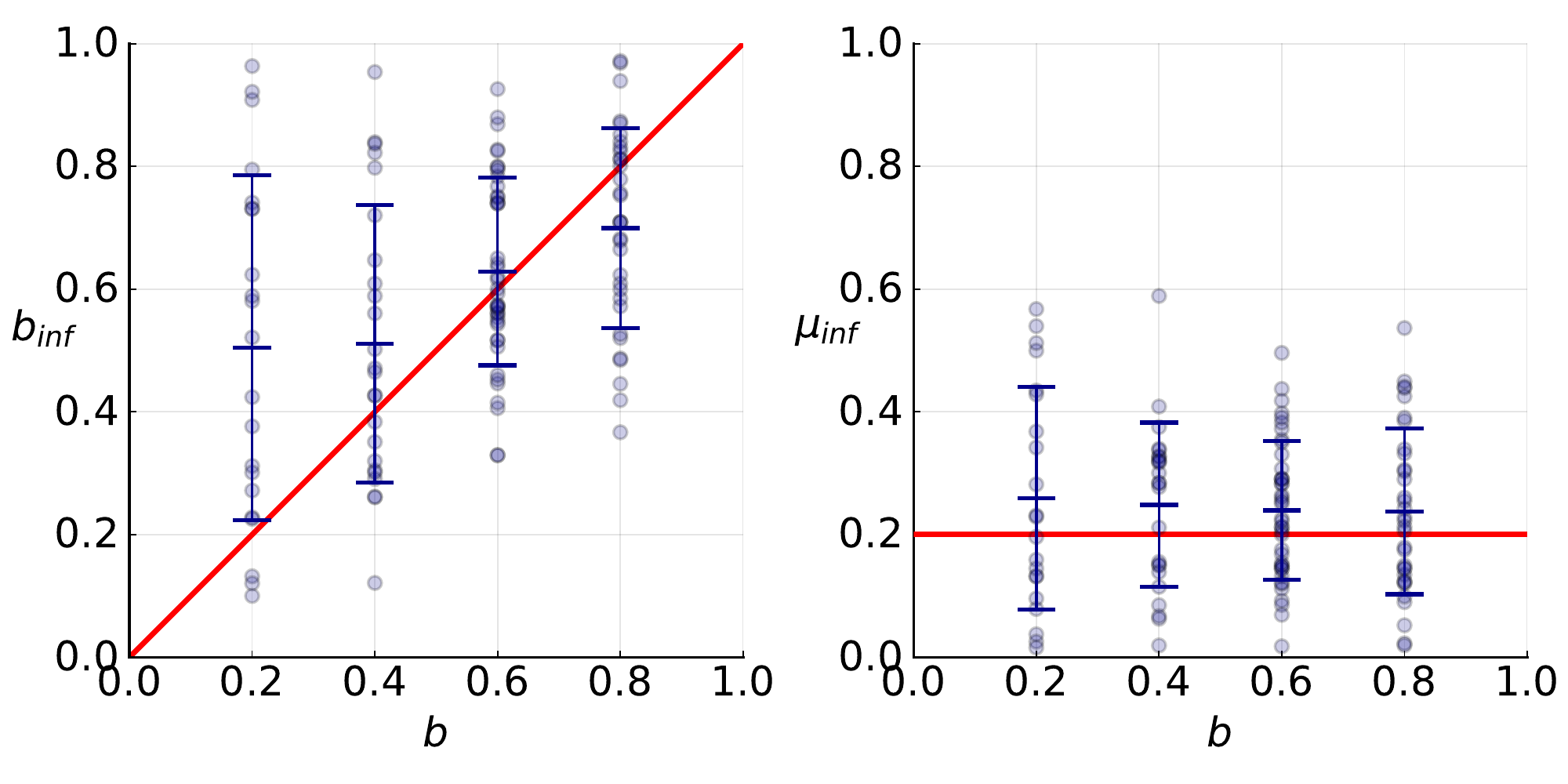}
    \caption{We use samples taken from spatial simulations (see text) to infer the death rate $b_{\textrm{\tiny{inf}}}$ and mutation rate $\mu_{\textrm{\tiny{inf}}}$ as in Fig.~\oldref{fig:inference2}. Figure symbols are as in Figures~\oldref{fig:inference1} and~\oldref{fig:inference2}. Finite sampling increases the variance of the inferred values, but inference is still possible provided the rate of cell death is sufficiently high, see text.
    }
    \label{fig:inference_spatial}
\end{figure}

To show that our method can be applied to current data with finite sampling, we apply a sampling scheme mimicking the one used in Ling et al.~\cite{ling2015extremely}, which gives whole-exome data on $270$ samples from a hepatocellular carcinoma. We used a cell-based simulation algorithm~\cite{drasdo2005single,metzcar2019review} to produce a population of $10^4$ cells with specific spatial positions, from which $200$ samples of about $50$ cells each were taken. We filtered out mutations below a frequency threshold of $1/200$ in all samples taken together, as well as mutations that occurred in a single sample only with a frequency below $1/3$.
For low mutation rates $\mu$ and death rates that were not too small, the inference results from sampled data compare well to the true underlying parameters $d/b$ and $\mu$, however with a higher variance than in the absence of sampling noise, see Figure~\oldref{fig:inference_spatial}. At relative death rate below $d/b \approx 0.2$, the inference of $d/b$ becomes poor, since clones become as likely to be lost through cell death as to be unobserved due to finite sampling.

The determination of the rates of birth, death, and mutation is a well-known problem in
different contexts. Experimentally, birth and death rates are accessible via live cell imaging~\cite{garvey2016high,hart2019high}, but
mutation rates are notoriously hard to estimate. The classic Luria-Dellbr\"uck experiment
allows to infer the mutation rate~\cite{luria1943mutations,kessler2013large,
kessler2015scaling}. However, it requires a large number of different populations with the same mutation rate. This makes the application to a single population evolving under potentially unique circumstances, like a tumour cell population or an epidemic, impossible~\cite{kendal1988pitfalls}.
A joint inference involving the rates of birth, death and mutation, based on the mutation frequency spectrum, has been given by Sottoriva and collaborators~\cite{williams2016identification}. In this approach, the quantity $\mu a/(a-b)$ has been inferred from bulk sequencing data across different tumour types. Using low-frequency mutations, it is in principle also possible to infer the mutation rate and the relative death rate separately~\cite{gunnarsson2021exact} from the frequency spectrum.
A separate inference of the mutation rate and the relative death rate has been achieved in
by Werner \textit{et al.}~\cite{werner2020measuring} using multi-region sequence data. Since their method is based on statistically distinguishing the number of mutations occurring during different numbers of cell divisions, it works best at a high expected number of mutations. Our approach operates in the complementary regime, where the expected number of mutations is small.

\section{Conclusions}
Cell division, death, and mutations lead to a constant genetic turnover: novel genetic variants enter a population and can be lost again. In a population of constant size, birth and loss of variants are on average exactly balanced. A growing population, on the other hand, supports a continuously increasing number of variants. However, this does not mean that no clones are lost and there is no turnover of genetic variants: fluctuations in the times at which cells divide and die can still lead to the extinction of clones, even when on average all clones grow in size. In this paper, we have set up a framework to quantify the effects of genetic drift in an exponentially growing population. To quantify the rate at which specific subpopulations (clones of a specific genotype or clades with a particular mutation) are lost from a population, we have defined two turnover parameters. The first one specifies the fraction of clones that have had their parent clone become extinct, and the second the fraction of clades that coincide with their parental clade. To calculate the turnover parameters analytically, we combined the stochastic dynamics of a birth-death process with first step analysis of the long-term fate of a subpopulation of cells.

The stochastic framework we have applied here can be used to calculated different observables in growing populations. A key property of the turnover parameter is that it can be computed from population data at one particular moment in time, rather than watching a population over time to see if a particular clone has become extinct. The result derived here can thus be useful to quantify the genetic turnover in a growing population observed at a particular time (for instance in the context of a tumour biopsy) to help infer the underlying rates of cell division, death, and mutation. Implementing this inference scheme requires a sufficient number of samples or cells from an individual tumour to reliably determine the turnover parameters even if low-frequency clones are lost due to sampling noise. Single-cell data or samples taken at a high spatial resolution~\cite{ling2015extremely} are promising starting points.

\ack{This work was supported by the DFG CRC 1310.}

\bigskip

\section*{References}


\end{document}